\documentclass[prb,showpacs,floatfix,superscriptaddress,amsmath,amssymb,twocolumn]{revtex4}
\usepackage{graphicx}% Include figure files
\usepackage{graphics}% Include figure files
\usepackage{dcolumn}% Align table columns on decimal point
\usepackage{bm}% bold math
\usepackage{amssymb}
\usepackage{hyperref}
\usepackage{color}

\usepackage{soul}

\newcommand{\ba}{\begin{eqnarray}}
\newcommand{\ea}{\end{eqnarray}}
\def\be{\begin{equation}}
\def\ee{\end{equation}}

\begin{document}

\title{A theoretical  model for tellurite-sulfates Na$_2$Cu$_5$(TeO$_3$)(SO$_4$)$_3$(OH)$_4$ and 
K$_2$Cu$_5$(TeO$_3$)(SO$_4$)$_3$(OH)$_4$}

% En orden alfabético?

\author{I. L. Bartolom\'e}
%\email{lamas@fisica.unlp.edu.ar}
\affiliation{IFLP - CONICET. Departamento de F\'isica, Facultad de Ciencias Exactas. Universidad Nacional de La Plata,C.C.\ 67, 1900 La Plata, Argentina.}

\author{L. Errico}
% \email{lamas@fisica.unlp.edu.ar}
\affiliation{IFLP - CONICET. Departamento de F\'isica, Facultad de Ciencias Exactas. Universidad Nacional de La Plata,C.C.\ 67, 1900 La Plata, Argentina.}
%\affiliation{Instituto de F\'isica La Plata (IFLP) y Departamento F\'isica, Facultad de Ciencias Exactas, Universidad Nacional de La Plata-CCT La Plata, CONICET, C.C. 67, CP 1900 La Plata, Argentina. }
\affiliation{Universidad Nacional del Noroeste de la Provincia de Buenos Aires (UNNOBA), Monteagudo 2772, CP 2700 Pergamino, Buenos Aires, Argentina}

\author{V. Fernandez}
%\email{lamas@fisica.unlp.edu.ar}
\affiliation{IFLP - CONICET. Departamento de F\'isica, Facultad de Ciencias Exactas. Universidad Nacional de La Plata,C.C.\ 67, 1900 La Plata, Argentina.}

\author{M. Matera}
\email{matera@fisica.unlp.edu.ar}
\affiliation{IFLP - CONICET. Departamento de F\'isica, Facultad de Ciencias Exactas. Universidad Nacional de La Plata,C.C.\ 67, 1900 La Plata, Argentina.}

\author{A.V. Gil Rebaza}
%\email{lamas@fisica.unlp.edu.ar}
\affiliation{IFLP - CONICET. Departamento de F\'isica, Facultad de Ciencias Exactas. Universidad Nacional de La Plata,C.C.\ 67, 1900 La Plata, Argentina.}

\author{ C.A.\ Lamas}
\email{lamas@fisica.unlp.edu.ar}
\affiliation{IFLP - CONICET. Departamento de F\'isica, Facultad de Ciencias Exactas. Universidad Nacional de La Plata,C.C.\ 67, 1900 La Plata, Argentina.}

\begin{abstract}
A theoretical model for two new tellurite-sulfates, namely Na$_2$Cu$_5$(TeO$_3$)(SO$_4$)$_3$(OH)$_4$ and 
K$_2$Cu$_5$(TeO$_3$)(SO$_4$)$_3$ (OH)$_4$ is determined to be compatible with ab-initio calculations.
The results obtained in this work show that some previous speculations in the literature about the couplings are correct, obtaining a model with a mixture of ferromagnetic and antiferromagnetic couplings.
We use a combination of numerical techniques to study the magnetic properties of the model. 
Our numerical calculations based on the density-matrix renormalization group method reveal that the system presents Ising-like 
magnetization plateaux at rational values  of the saturation magnetization.

%The model obtained captures the observed magnetic characteristics.

\end{abstract}

\maketitle

\section{Introduction}

Recently, Yingying Tang et al.\cite{experimental_kagome_strip} synthesized  by
hydrothermal reaction, two new tellurite-sulfates (TS) with a distorted Kagom\'e strip structure:
Na$_2$Cu$_5$(TeO$_3$)(SO$_4$)$_3$(OH)$_4$ and  K$_2$Cu$_5$(TeO$_3$)(SO$_4$)$_3$(OH)$_4$ (Na-TS and K-TS in the following).
In both compounds, the magnetic behavior is associated with the single unpaired electron associated with each ${\rm Cu}^{+2}$ ions, localized over a 1D kagomé strip sub-lattice.
This particular geometry corresponds to the one dimensional version of the paradigmatic two dimensional Kagom\'e lattice for which some experimental  realizations for $S=1/2$ as the Herbertsmithite ZnCu$_3$(OH)$_6$Cl$_2$ \cite{K-exp-1}, the $\alpha$-vesignieite
 BaCu$_3$V$_2$O$_8$(OH)$_2$\cite{K-exp-2},  and [NH$_4$]$_2$ [C$_7$H$_{14}$N][V$_7$O$_6$F$_18$]$_5$ \cite{K-exp-3} were estudied.

\begin{figure}[t!]
\includegraphics[width=0.79\columnwidth]{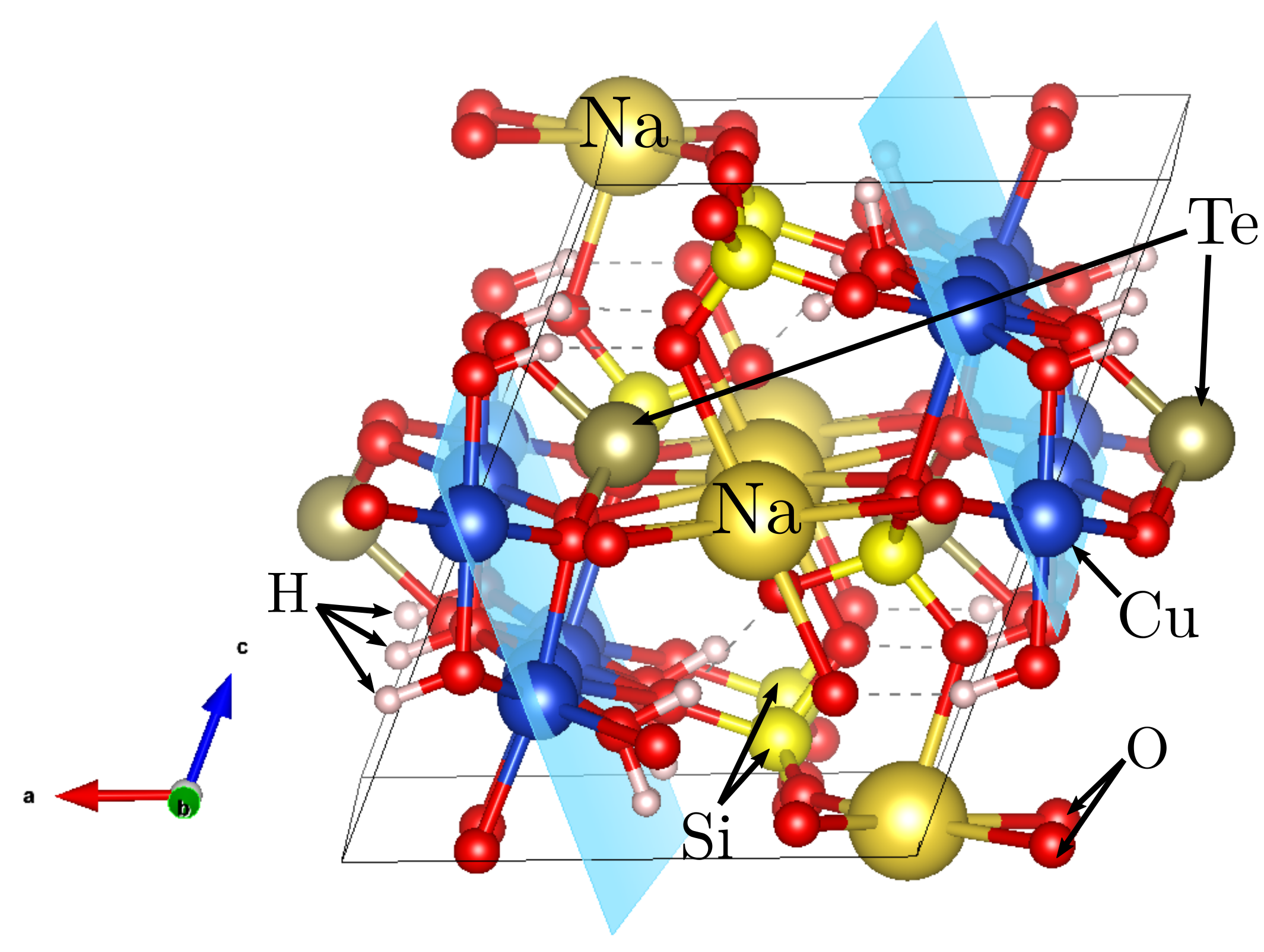}
\caption{Crystal structure corresponding to Na$_2$Cu$_5$(TeO$_3$)(SO$_4$)$_3$(OH)$_4$(Na-TS). The compound  K$_2$Cu$_5$(TeO$_3$)(SO$_4$)$_3$(OH)$_4$ is isostructural with Na-TS. }
\label{fig:structure}
\end{figure}

The crystal structure of the compounds is schematized in Fig. \ref{fig:structure} and the simplified magnetic geometry we consider is shown in Fig. \ref{fig:2Dlattice}.  
We show that several magnetic properties like magnetic plateaux are determined by the geometry of the plaquette.

 The synthesis of these compounds has aroused great interest in the study of the magnetic phase diagram of models with this Kagomé strip geometry\cite{Schulenburg2002,Acevedo2019,morita2018magnetization,Acevedo2020,Morita2021,morita2021resonating,strip-dmrg,strip-dmrg-comment,strip-dmrg-reply,strip-lecheminant}. In this sense, the presence of magnetization plateaux\cite{plateaux_kagome_strip_2021}, a Haldane-like phase\cite{Morita2021} and localized magnon crystal phases have been detected\cite{Acevedo2019,Okuma2019}.
The studies carried out so far describe general phase diagrams in a parameter space that, a priori, is not related to the couplings that describe these materials.
Improving the theoretical description then requires estimating the coupling constants of the effective magnetic model. As proposed by Noodelman \cite{noodleman1981}, a way to determine these coupling constants is by comparing the spectrum of the reduced model to those obtained by setting the corresponding magnetic configurations in Density Functional Theory (DFT) based calculations. The original method was successfully applied in the literature to compute the magnetic coupling constants of many compounds.
However, as the number of coupling constants and atoms in the supercell grows, the direct application of the method becomes challenging: since the number of possible magnetic configurations grows exponentially with the number of magnetic atoms,  and the evaluation of the energy of each configuration is computationally expensive, to exhaust the full set of magnetic configurations becomes impractical even for a small number of magnetic atoms. On the other hand, choosing a small set of magnetic configurations could introduce a large bias in the determination of the coupling constants.
To overcome these issues, a novel strategy based on Noodelman's breaking symmetry method was proposed\cite{matera_espectro}. In this work, that methodology is used to determine the couplings in the magnetic model describing the tellurite-sulfates.
A discussion about this coupling determination is presented and we show that the $S=1/2$ Heisenberg model with these couplings describes the magnetic properties
of the system and allows to determine qualitatively the behavior of the magnetic transitions.

Inspired by the experimental determination of the atomic distance we propose a model with five different magnetic couplings and determine the set of couplings values compatible with the energies calculated by density functional methods.
The resulting model is numerically studied, determining the zero temperature magnetization curve by density-matrix renormalization group (DMRG) calculations. We also determine some thermodynamical quantities for small systems by exact diagonalization.

We analyze the magnetic plateaux at zero temperature in the context of the Oshikawa-Yamanaka-Affleck (OYA) theorem\cite{OYA} which provides the necessary condition for the existence of these magnetization plateaux as $$ NS(1-m)=\hbox{integer}$$
where $N$ is the number of spins in the ground state (G. S.) cell presenting spatial periodicity and $m=M/M_{sat}$ is the normalized magnetization per site.
If the translational symmetry in the G. S. is preserved, then $N=5$ and the magnetization curve may have plateaux at $m=1/5$ and $m=3/5$.
In the following, we show that the G. S. periodicity is enlarged to $N=10$, but still only the semi-classical 
plateux at $m=1/5$ and $m=3/5$ are present.

The paper is organized as follows; in Sec. \ref{sec:magneticmodel} basic properties of the lattice and magnetic degrees of freedom are discussed.
In Sec. \ref{sec:ab-initio}  we discuss details of the couplings estimation by following the methodology of ref \cite{matera_espectro}. Estimated values of the coupling constants are also presented.
Then, in Sec. \ref{sec:magbehaviour}, we study the magnetic properties arising from the fitted model, both for large systems in the zero-temperature limit, by DMRG calculations, and at finite temperature, through full diagonalization of the quantum model for small systems. 
Finally in Sec. \ref{sec:conclusions}, we present the conclusions and perspectives.

\begin{table}[t!]
\centering
\begin{tabular}{|c|c|c|}
\hline
                       &  Na-TS     &   K-TS     \\ \hline
 $a$ ($\AA$)           &  7.294(3)  &  7.467(6)  \\ 
 $b$ ($\AA$)           & 12.005(4)  & 12.177(9)  \\ 
 $c$ ($\AA$)           &  9.214(3)  &  9.397(4)  \\ 
 $\alpha$  	       &  90.0      &  90.0      \\ 
 $\beta$  	       & 111.160(6) & 111.352(8) \\ 
  $\gamma$  	       &  90.0      &  90.0      \\ \hline
\end{tabular}
\caption{\label{tab:structural}Crystal Structural parameters for Na-TS and K-TS compounds.}
\end{table}

\section{The magnetic model}
\label{sec:magneticmodel}

Na-TS and K-TS are isostructural compounds that crystallize in a monoclinic structure with 
space group P2$_1$/m, see Figure \ref{fig:structure}.  The structural information for both compounds is reported in Table I. Atomic positions of each atom in the 
structure for both Na-TS and K-TS can be found in Ref. \cite{experimental_kagome_strip}. In its three inequivalent crystallographic 
sites, the ${\rm Cu}^{+2}$ ions form distorted CuO$_6$ octahedral with bond lengths ranging from $1.85$ to $2.31$ $\AA$ (Na-TS)
and $1.88$ to $2.50$ $\AA$ (K-TS),  exhibiting a Kagom\'e-strip arrangement which can be regarded as a 
dimensional reduction of Kagom\'e-lattice. 

\begin{figure}[t!]
\includegraphics[width=0.79\columnwidth]{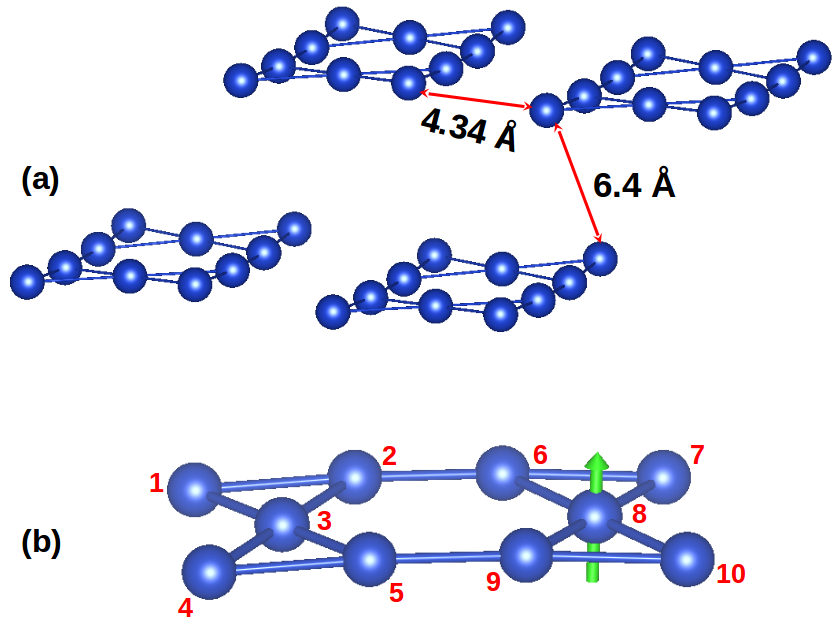}
\caption{a) Distance between Kagom\'e-strip lattice for Na-TS. b) labels of the Cu atoms in the 2D Kagom\'e-strip.}
\label{fig:2Dlattice}
\end{figure}
\begin{table}[t!]
\centering
\begin{tabular}{|c|c|c|c|}
\hline
                     &  Na-TS     &   K-TS  & Coupling constant   \\ \hline
 $d_{Cu2-Cu6}$ ($\AA$) &  3.01      &  3.07   &  $J_d$   \\
 $d_{Cu5-Cu9}$ ($\AA$) &  2.84      &  2.89   &  $J_u$ \\
 $d_{Cu1-Cu2}$ ($\AA$) &  3.08      &  3.11   &  $J_0$ \\
  $d_{Cu1-Cu3}$ ($\AA$) &  3.07      &  3.11   &  $J_1$ \\
  $d_{Cu2-Cu3}$ ($\AA$) &  2.94      &  2.96   &  $J_2$ \\  \hline
\end{tabular}
\caption{\label{tab:structural2}Distances between Cu atoms corresponding to each coupling constant.}
\end{table}

Our aim in this work is to study the magnetic behavior of Na-TS and K-TS. Since the Te, S, O, H, and Na/K ions do not present spin polarization,
the spin-lattice is determined by the ${\rm Cu}^{+2}$ ions that form the Kagom\'e-strip lattice, as can be seen in Figure \ref{fig:2Dlattice}. Since each
magnetic ion has a single localized unpaired electron, its magnetic degree of freedom can be described as a spin  $S=1/2$.

In order to build a simple effective model for the magnetic degrees of freedom, we propose a symmetric Heisenberg model
\ba
H=-\sum_{(i,j)} J_{i,j} \vec{\bf{S}}_{i}\cdot \vec{\bf{S}}_{j}.
\ea
with $\vec{\mathbf{S}_{i}}=\frac{1}{2}(\sigma_{x,i}, \sigma_{y,i}, \sigma_{z,i})$ the spin vector, and $J_{i,j}$ the coupling constants. To  determine them, we impose the constraint that the difference between the DFT energy and the energy of the Heisenberg model for a given set of couplings must be lower than the DFT energy error ($1 {\rm mRy}$). This is our compatibility criterium.

\begin{figure}[t!]
\includegraphics[width=0.7\columnwidth]{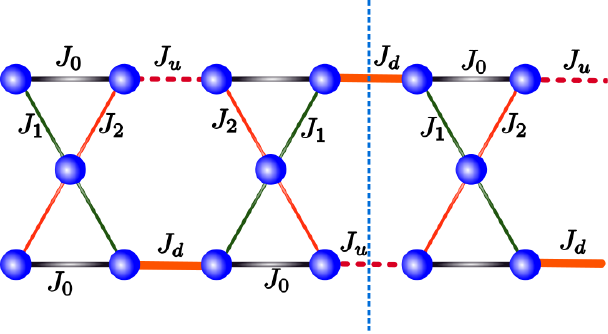}
\caption{Coupling constants in the Heisenberg model. Notice that the inversion symmetry around the dashed-light-blue  line reported in the crystallographic data  implies an alternated $J_1$-$J_2$ pattern in the diagonal bonds. }
\label{fig:Jotas}
\end{figure}

The Kagom\'e-strip lattice formed by the ${\rm Cu}$ ions in the Na-TS and K-TS compounds is quite distorted, showing five different Cu-Cu bond lengths (see Table \ref{tab:structural2}). 
The nearest ${\rm Cu}-{\rm Cu}$ distance between Kagom\'e-strips is in the order of 4.34 $\AA$ (Na-TS) and 4.44 $\AA$ (K-TS), respectively, while the 
shortest distance between the layers are 6.4 $\AA$ (Na-TS) and 6.7 $\AA$ (K-TS), respectively see Figure \ref{fig:2Dlattice}a. \\

In the present work, we considered interactions up to 3.1 $\AA$, Figure \ref{fig:2Dlattice}b, i.e. we discard the interactions between contiguous layers 
and neighbors strips. This implies the calculation 
of five exchange couplings, schematized in Fig. \ref{fig:Jotas}, ($J_0$, $J_1$, $J_2$, $J_u$, $J_d$) of an effective spin$-\frac{1}{2}$ Heisenberg model. 

\begin{figure}[b!]
  \centering
  \scalebox{.6}{\includegraphics{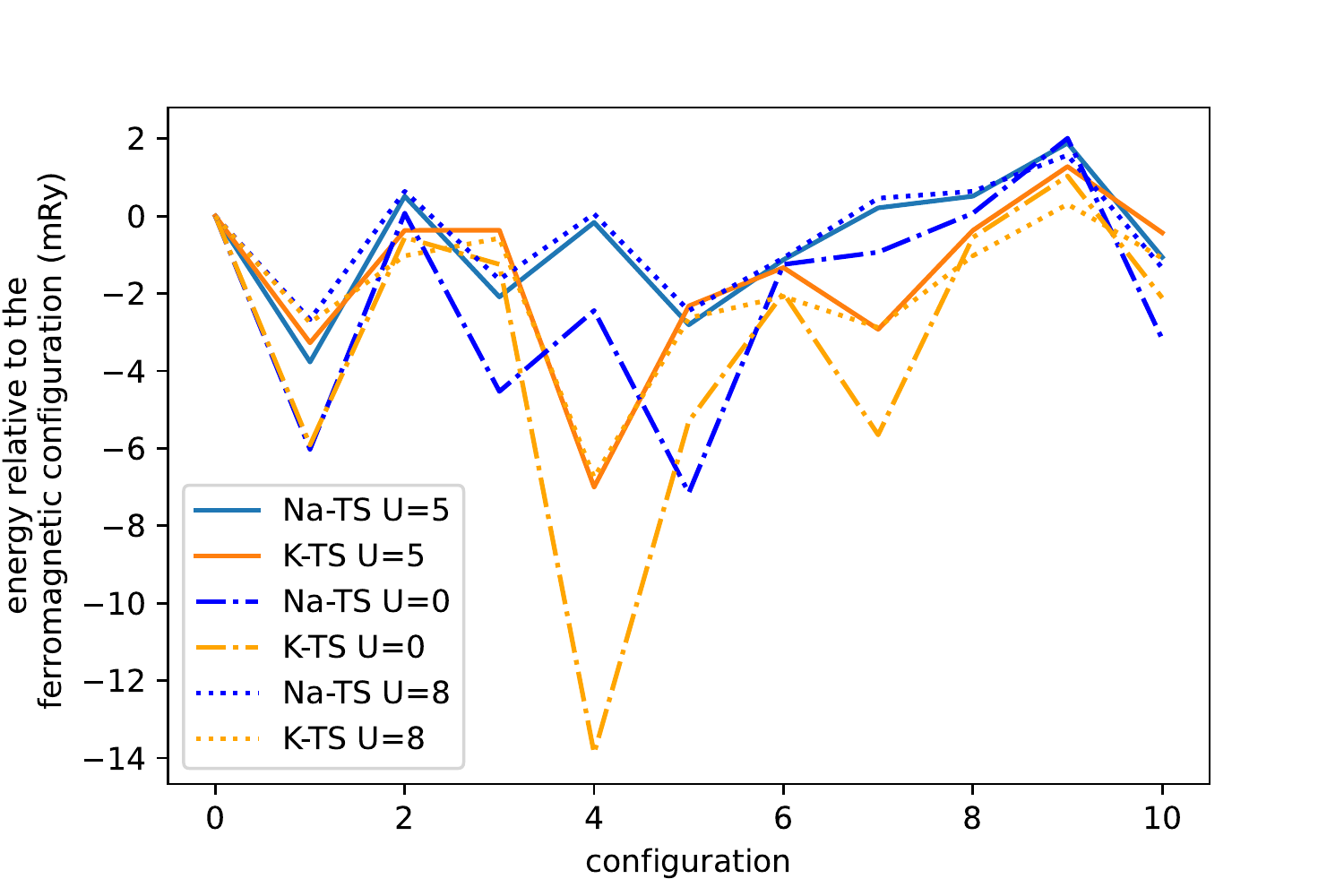}}
  \caption{\label{fig:energies}DFT estimated energies for the different spin configurations described in Table \ref{tab:arrowsMM} for both compounds and different values of the
    Hubbard's parameter $U$. Notice that for $U=5\;{\rm eV}$ and $U=8\;{\rm eV}$ the estimated energies are not significantly different.}  
\end{figure}

Once the model was established, DFT-based first-principles calculations were performed to determine the total energy of different spin configurations of Na-TS and K-TS. Then, these configurations were 
mapped to an appropriate spin model to obtain the exchange couplings. To obtain accurate $J_{n}$ values and their error bars, the methodology proposed in \cite{matera_espectro} was followed. The energies calculated via DFT are presented in Table \ref{tab:arrowsMM} and Fig. \ref{fig:energies}.

\section{Fitting coupling constans from ab-initio simulations}
\label{sec:ab-initio}

Following the procedure previously described, we start by choosing a set of eleven magnetic configurations over a supercell of dimensions $2a\times 2b \times c$ (see Table \ref{tab:arrowsMM}). 
The selection of which configurations (and how many of them must be taken into account) was made in order to optimize the sensitivity of the energy associated with the configurations for the  Heisenberg model under small changes in the coupling constants\cite{matera_espectro}.

\begin{table}[t!]
\centering
\begin{tabular}{|c|c|c|c|}
\hline
               &  Magnetic moment alignment & $E_{Na-TS}$ & $E_{K-TS}$ \\ \hline
  $|0\rangle$   &  $|\uparrow\;\;\uparrow\;\;\uparrow\;\;\uparrow\;\;\uparrow\;\;\uparrow\;\;\uparrow\;\;\uparrow\;\;\uparrow\;\;\uparrow\; \rangle$
                                            &0& 0  \\ 
  $|1\rangle$   &  $|\uparrow\;\;\uparrow\;\;\downarrow\;\;\uparrow\;\;\uparrow\;\;\uparrow\;\;\uparrow\;\;\downarrow\;\;\uparrow\;\;\uparrow\; \rangle$
                                            &-3.77& -3.21   \\ 
  $|2\rangle$   &  $|\downarrow\;\;\uparrow\;\;\downarrow\;\;\uparrow\;\;\downarrow\;\;\downarrow\;\;\uparrow\;\;\downarrow\;\;\uparrow\;\;\downarrow\; \rangle$
                                            &0.51&   -0.37  \\ 
  $|3\rangle$   &  $|\uparrow\;\;\uparrow\;\;\downarrow\;\;\downarrow\;\;\downarrow\;\;\uparrow\;\;\downarrow\;\;\downarrow\;\;\downarrow\;\;\downarrow\; \rangle$
                                            &-2.09&  -0.37  \\ 
  $|4\rangle$   &  $|\uparrow\;\;\downarrow\;\;\downarrow\;\;\downarrow\;\;\uparrow\;\;\downarrow\;\;\uparrow\;\;\downarrow\;\;\uparrow\;\;\uparrow\; \rangle$
                                            &-0.16&   -7.00  \\ 
  $|5\rangle$   &  $|\downarrow\;\;\uparrow\;\;\downarrow\;\;\uparrow\;\;\downarrow\;\;\uparrow\;\;\downarrow\;\;\uparrow\;\;\downarrow\;\;\downarrow\; \rangle$
                                      &-2.81      &  -2.32 \\
  $|6\rangle$   &  $|\uparrow\;\;\downarrow\;\;\downarrow\;\;\uparrow\;\;\downarrow\;\;\uparrow\;\;\downarrow\;\;\downarrow\;\;\uparrow\;\;\downarrow\; \rangle$
                                      &-1.13     &  -1.33 \\
  $|7\rangle$   &  $|\uparrow\;\;\downarrow\;\;\downarrow\;\;\downarrow\;\;\downarrow\;\;\uparrow\;\;\uparrow\;\;\downarrow\;\;\downarrow\;\;\downarrow\; \rangle$
                                      & 0.21     &  -2.92  \\
  $|8\rangle$   &  $|\uparrow\;\;\downarrow\;\;\downarrow\;\;\downarrow\;\;\uparrow\;\;\uparrow\;\;\downarrow\;\;\downarrow\;\;\downarrow\;\;\uparrow\; \rangle$
                                      & 0.51     &  -0.37  \\
  $|9\rangle$  &  $|\downarrow\;\;\downarrow\;\;\downarrow\;\;\uparrow\;\;\uparrow\;\;\uparrow\;\;\uparrow\;\;\downarrow\;\;\downarrow\;\;\downarrow\; \rangle$
                                       &1.88     &  1.28 \\
  $|10\rangle$  &  $|\uparrow\;\;\uparrow\;\;\downarrow\;\;\uparrow\;\;\downarrow\;\;\downarrow\;\;\downarrow\;\;\uparrow\;\;\downarrow\;\;\downarrow\; \rangle$
                                     &-1.06       &  -0.43  \\ \hline
\end{tabular}
\caption{\label{tab:arrowsMM}Subset of magnetic configurations of the Cu atoms obtained with the algorithm presented in \cite{matera_espectro}, and the corresponding energies obtained from DFT simulations. Arrows indicate the relative magnetic moment orientation of each Cu ions of the Figure\ref{fig:2Dlattice}b represented by $|{\rm Cu1}\;{\rm Cu2}\;\cdots\;{\rm Cu10}\; \rangle$. The values of the energies, in ${\rm mRy}$ are relative to the corresponding ferromagnetic configuration and obtained for calculations with $U=5\;{\rm eV}$. A comparison with the results obtanied for other values of $U$ are depicted in Figure \ref{fig:energies}.}
\end{table}

%%%%

\subsection{Determination of the magnetic configuration energies}

%%%%%

To determine the energy of each configuration in the effective Heisenberg model, we need an estimation of these energies for the full electronic model. To evaluate them,  first-principles calculations were performed in the framework of the DFT \cite{DFT},
where the self-consistent Kohn-Sham equations have been solved using the pseudopotential and plane-wave method as implemented in the Quantum Espresso code \cite{quantum_espresso}, where the core ions were described by Ultrasoft pseudopotential (USPP) from the Standard Solid State Pseudopotentials (SSSP) repository \cite{prandini2018precision}. 
The exchange-correlation part was described by the Perder-Burke-Ernzerhof parametrization of the Generalized Gradient Approximation (PBE-GGA) \cite{perdew1996generalized}. 
The kinetic energy cutoff for the wave function and charge-density used were 80 and 800 Ry, respectively. The reciprocal space was described by a 
dense mesh-grid of $12\times 12\times 12$ k-points and the spin-polarized effect was considered to explore different collinear magnetic states. 
In order to enhance the electronic structure description of the systems under study, we have added the Hubbard term (DFT+U) using the 
simplified rotationally invariant formulation \cite{cococcioni2005linear}. The value of $U = 5.0\;  \rm{eV}$ for the $3d$-Cu orbitals was obtained using the linear-response 
approach based on the Density Functional Perturbed Theory (DFPT) \cite{cococcioni2005linear}.\\
For each spin configuration, lattice parameters and angles were fixed at the experimental ones but all the atomic positions were 
relaxed to minimize the forces on the ions. Structural optimizations were performed until achieve that these forces were below the 
tolerance criteria $|\nabla E| \leq 0.1\,\rm{eV}/\AA$. The obtained atomic positions and bond-lengths are nearly independent of the spin configuration considered and are in excellent agreement with the experimentally determined ones\cite{experimental_kagome_strip}.

From DFT+$U$ calculations, for $U=5\,\rm{eV}$, and irrespective of the magnetic configuration considered the absolute value of the magnetic moment of the Cu atoms  are $2.02\;\mu_B$ and $2.07\;\mu_B$ for Na-TS and K-TS, respectively. These results for the magnitude of the magnetic moments at the Cu sites are in excellent agreement with the experimental ones \cite{experimental_kagome_strip}. The direction of the magnetic moments is perpendicular to the 2D Cu-plane, i.e., the strip-plane (see Fig.\ref{fig:2Dlattice}).

\subsection{Coupling constants estimation}
\label{sec:exchange-couplings}

With the energies estimated at each fixed value of $U$, a first estimation of the set of coupling constants was obtained  by means of a least-square fitting of the energies predicted in the Heisenberg model.
A sample of 10000 candidates for the set of coupling constants, with a gaussian distribution around the fitted values, was generated afterward.  From these samples,  those whose configurational energy differed from the DFT estimation in a difference larger than the DFT estimated error (around $1 \rm{mRy}$) were discarded.  The width of the gaussian distribution was chosen in a way that around $50\%$ of the samples resulted rejected. Then, the value for each coupling constant was assumed to lay between the minimal and maximal values reached over the remaining configurations. 
For the compound Na-TS (K-TS),  the estimation for the coupling constants with their corresponding uncertainties, assuming different values of the $U$ constant,  are reported in Table  \ref{tab:jotasNa} (\ref{tab:jotasK}).
Note that although varying the value of $U$ changes the general scale of couplings, its relative value and sign remains invariant, so the conclusions obtained about the magnetic behavior do not depend on the value of $U$.
\begin{table}[b!]
\centering
\begin{tabular}{|c|c|c|c|c|}
\hline
  & $U=2{\rm eV}$ & $U=5{\rm eV}$ & $U=8{\rm eV}$ \\ \hline
  $E_0$ [$\rm{Ry}$]   &      $-5582.490(1)$  &     $-5582.175(1)$ &       $-5581.871(1)$\\ \hline
  $W$ [$\rm{mRy}$]   &      $0.855$  &     $0.691$ &       $0.543$\\ \hline
 $J_0/W$     &       $-0.3(2)$  &  $-0.3(1)$ &    $-0.3(3)$ \\ \hline
 $J_1/W$      &      $0.2(3)$  &   $0.2(1)$  &    $0.3(6)$	\\ \hline
 $J_2/W$   	 &   $-0.9(2)$   & $-0.9(2)$   &  $-1.0(5)$	\\ \hline
$J_u/W$   	 &   $1.0(5)$   &  $1.0(3)$    &  $0.8(9)$\\ \hline
$J_d/W$   	 &   $0.4(5)$   &  $0.2(3)$     & $0.3(8)$\\ \hline
$T_{CW}$[K]         &       $-8.10$          &   $-8.74$           &    $-7.72$\\ \hline
\end{tabular}
\caption{\label{tab:jotasNa}Couplings constants compatible with the DFT energies for Na-TS. $W$ is the magnitude of the strongest coupling constant ($\max_{i} |J_i|/W=1$).}
\end{table}
\begin{table}[b!]
\centering
\begin{tabular}{|c|c|c|c|c|}
\hline
   & $U=2{\rm eV}$ & $U=5{\rm eV}$ & $U=8{\rm eV}$ \\ \hline
  $E_0$ [$\rm{Ry}$]   &  $-5597.510(1)$      & $-5597.194(1)$      & $5596.890(1)$  \\ \hline
  $W$ [$\rm{mRy}$]   &      $0.871$  &     $0.599$ &       $0.555$\\ \hline
  $J_0/W$      &   $-0.3(2)$  & $-0.4(1)$   & $-0.4(3)$ \\ \hline
  $J_1/W$      &   $0.3(2)$  &  $0.3(5)$   & $0.4(5)$	\\ \hline
 $J_2/W$      &  $-1.0(2)$   & $-1.0(6)$   &  $-1.0(6)$	\\ \hline
 $J_u/W$      &   $0.9(3)$   &  $0.9(8)$   & $0.8(9)$\\ \hline
 $J_d/W$     &   $0.3(3)$   &  $0.3(8)$   &  $0.(1)$\\ \hline
$T_{CW}/W$   &     $-11.01$         &     $-10.90$         &  $-10.52$   \\ \hline
\end{tabular}
\caption{\label{tab:jotasK}Couplings constants compatible with the DFT energies for K-TS. $W$ is the magnitude of the strongest coupling constant ($\max_{i} |J_i|/W=1$).}
\end{table}
%
%  
%
% The results for the $J_n$ exchange couplings corresponding to Na-TS and K-TS as a function of the $U$ values are presented in Table \ref{tab:jotasNa} and Table \ref{tab:jotasK} respectively. 
%
%Again, notice that the relative values, as well as the signs are independent of the $U$ parameter, which affects just the overall factor in a ${\cal O}(1)$ factor.
Moreover, the estimated signs in the couplings are consistent with the experimental observations, and with the estimations provided by the Goodenough's rules \cite{goodenough}.

%%%%%%%%%%%%%%%%%%%%%%%%%%%%%%%%%%%%%%%%%%%%%%%%%%%%%%%%%%%%%%%%%%%%%%%%%%%%%%%%%%%%%%

\section{Magnetic behavior of the model}
\label{sec:magbehaviour}

\subsection{Curie-Weis temperature}
\label{sec:CW}

Once the couplings of the magnetic model have been determined and found to be reliable with respect to different values of $U$ and consistent with experimental evidence, we explore the behavior of the effective magnetic model. From the estimated coupling constants,
and through a mean field approximation for large $T$, the value of the Curie-Weiss temperature can be estimated as
\ba
\Theta_{CW} = \frac{1}{N} \left(\frac{S(S+1)}{3/2}\right)\frac{\sum_{(ij)} J_{(ij)}}{k}
\label{eq:curie-temp-mf}
\ea
where $N$ is the number of  spins  in the unit cell, $k\approx 0.08617 \; \rm{meV}/{\rm K}$ is the Boltzmann constant, and the summation cover all the connected pair of spins.
For $S=1/2$ we have
$$
\Theta_{CW} = \frac{\sum_{(ij)} J_{(ij)}}{2 N k}
$$
From this expression and the fitted values of the coupling constants, the model predicts $\theta_{CW}$(Na-TS) $= -8.73 {\rm K}$, $\theta$(K-TS) $= -9.46 {\rm K}$ which are in good agreement with the experimental ones ( $\theta$(Na-TS)$ = -6.1(8) {\rm K}$ and $\theta$(K-TS)$ = -13.9(4) {\rm K}$)\cite{experimental_kagome_strip}. These negative Weiss temperatures indicate a global antiferromagnetic behavior of both systems, also consistent with the experimental observations. We notice again that the absolute value of these temperatures does not appreciably vary with the choice of the Hubbard's $U$ parameter, so the estimation for Curie-Weis temperature is robust.

\subsection{Magnetization at zero temperature}
\label{sec:dmrg}

In order to study the magnetization as a function of the applied magnetic field
we use DMRG calculations for large stripes ($120$ spins).
For the calculations, we kept up to $500$ states throughout the work, 
which showed to be enough to achieve a good precision.
We calculate the g. s.  energy corresponding to each magnetization sector and determine the magnetization as a 
function of the applied magnetic field for both materials. In the following, we present results for Na-TS 
since the results corresponding to K-TS are similar.  

\begin{figure}[t!]
\includegraphics[width=0.99\columnwidth]{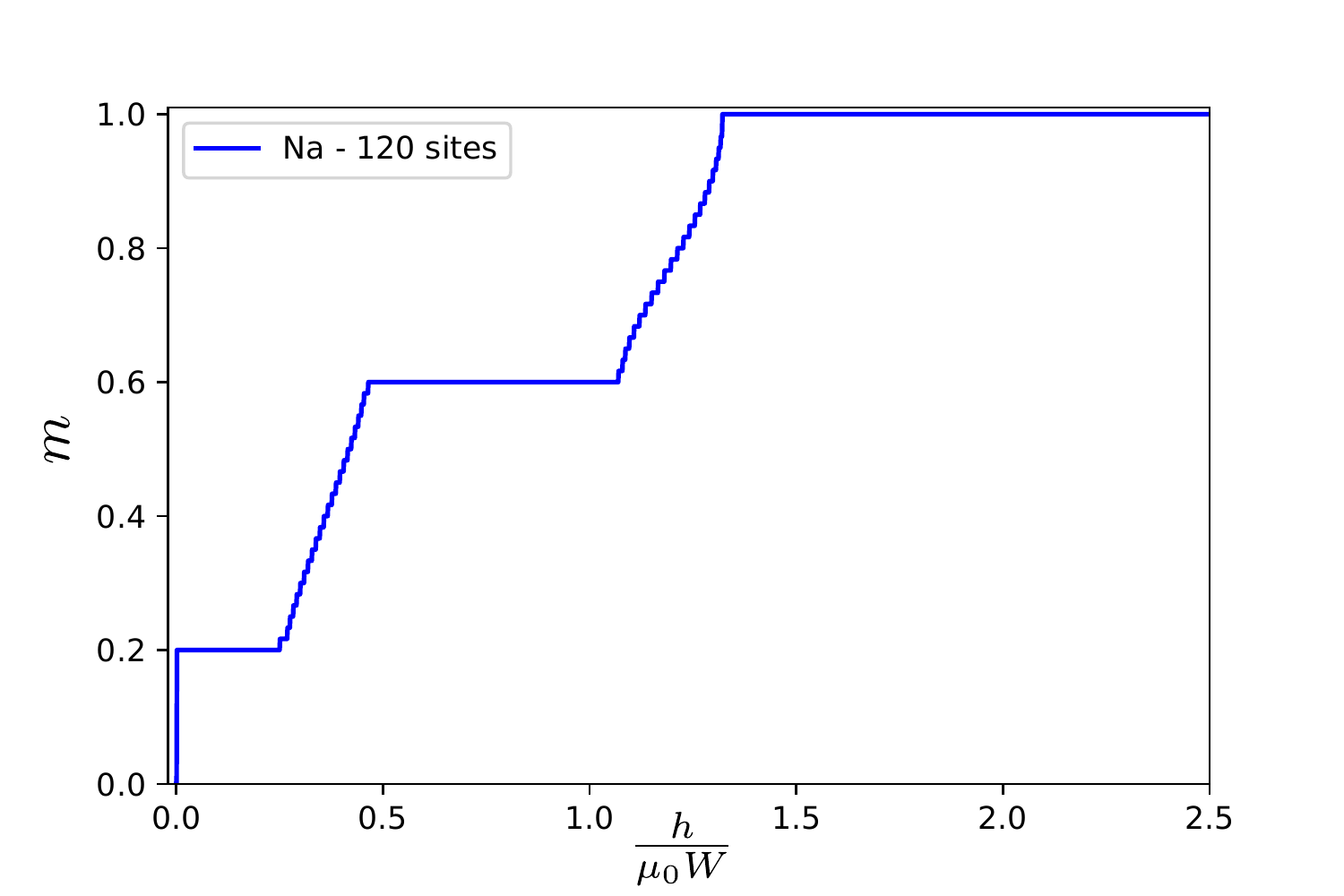}
\caption{Magnetization vs magnetic field corresponding to the Heisenberg model with the couplings estimated for Na-TS 
by ab-initio calculations with $U=5$ eV (see Table \ref{tab:jotasNa}). Semi-classical plateaux at $m=M/M_{sat}=1/5$ and $m=M/M_{sat}=3/5$ are clearly observed. }
\label{fig:mvsh_Na}
\end{figure}

\begin{figure}[t!]
\includegraphics[width=0.99\columnwidth]{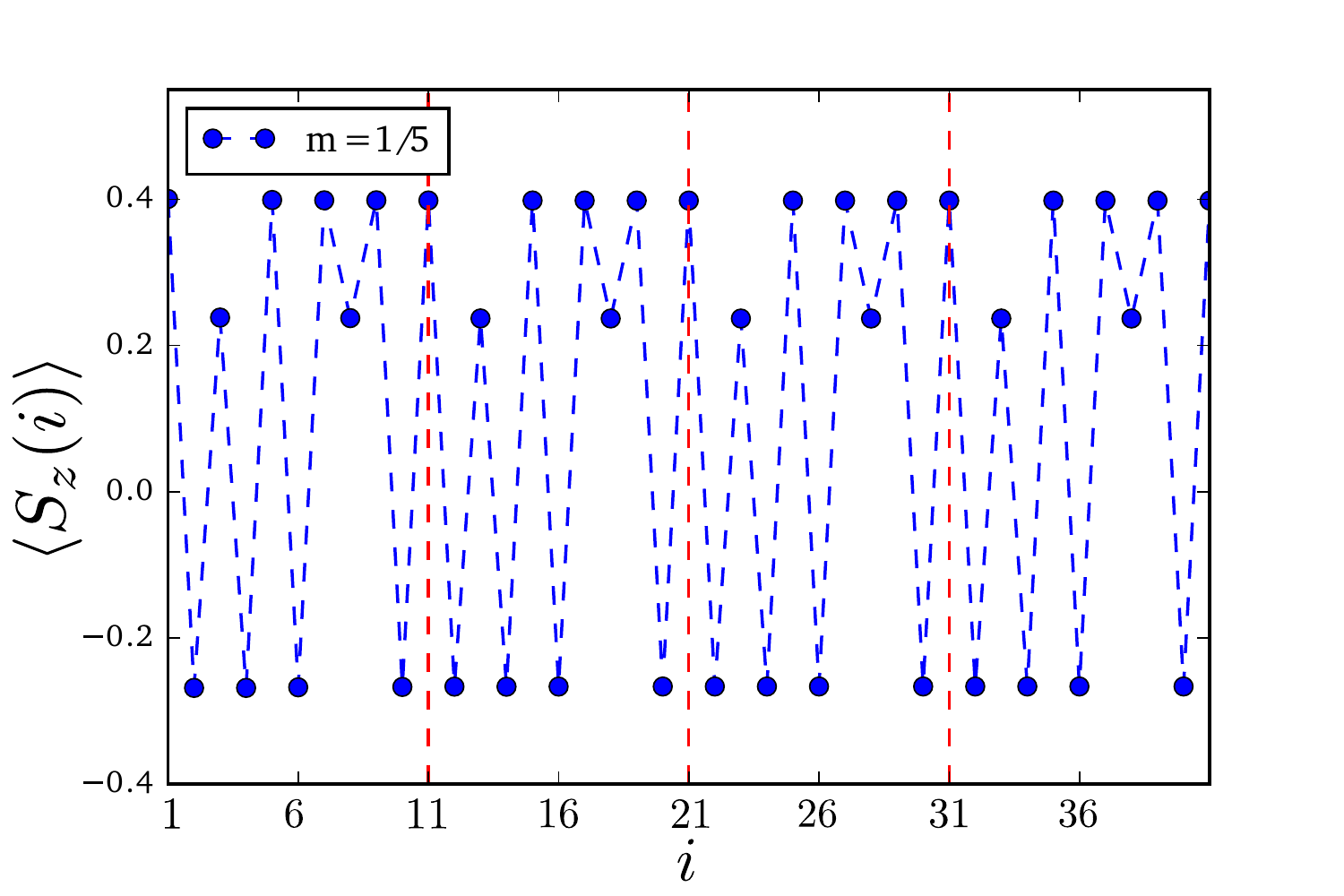}
\caption{Magnetization profile at $m=1/5$ as a function of the spin position corresponding to the Heisenberg model with the couplings estimated for Na-TS by DFT calculations.  }
\label{fig:sz_dmrg}
\end{figure}

The results for the magnetization vs magnetic field corresponding to Na-TS are showed 
in Fig. \ref{fig:mvsh_Na}. 
We observe the presence of the magnetic plateaux at $m=1/5$ and $m=3/5$. These magnetic plateaux are
allowed by the OYA criterium\cite{OYA}. 
In Fig. \ref{fig:sz_dmrg} we show $S^{z}_{i}$ value corresponding to the g. s.  at $m=1/5$ as a function of the site label.
The observed magnetic profile is consistent with a semi-classical plateau similar to that we could expect for an Ising model. 
For this magnetization sector, the g. s.  periodicity is $N=10$ and the same periodicity is observed for $m=3/5$.
The OYA criterium represents a necessary (but not sufficient) condition for the appearance of magnetic plateaux.
As in both materials, the observed G. S. periodicity is $N=10$ sites, the OYA criterium allows to have plateaux 
at $m=0$, $m=1/5$, $m=2/5$, $m=3/5$, $m=4/5$. It is interesting that only the semi-classical plateaux at $m=1/5$ and $m=3/5$ are present.

\begin{figure}[t!]
\includegraphics[width=0.69\columnwidth]{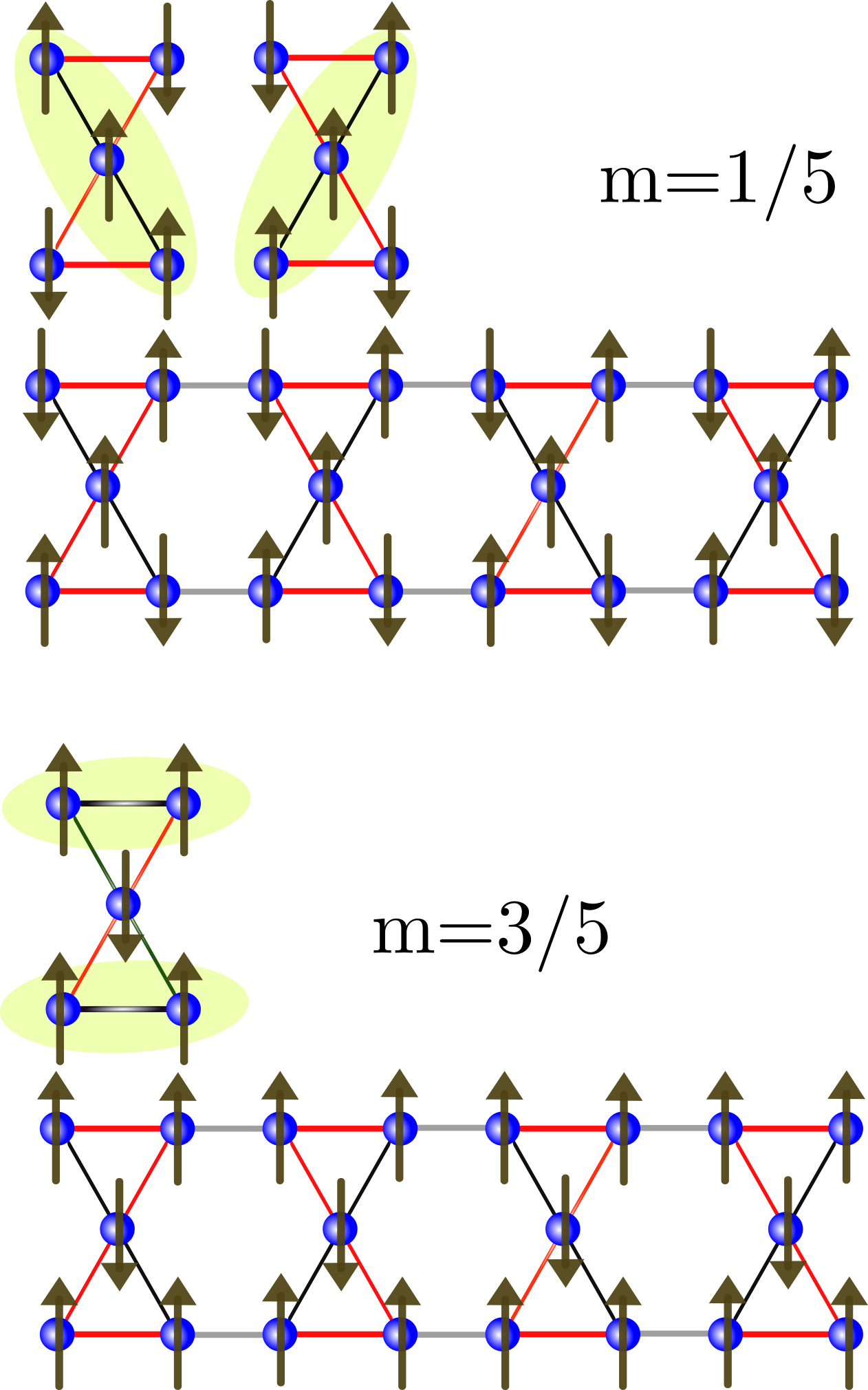}
\caption{Schematic representation of the expected Ising-like  configurations at $m=1/5$ and $m=3/5$. 
Top: two minimal energy configurations for $m=1/5$.
Bottom: Ising-like configuration for $m=3/5$}
\label{fig:ising-picture}
\end{figure}

In order to describe from a classical perspective the magnetic profile observed at $m=1/5$ and $m=3/5$,
let us first consider the simplest Ising limit of an isolated 5 sites plaquette ($\vec{S}_{j}=(0,0,S_j^z)$). 
The exchange terms in the Hamiltonian for the plaquette reads
\begin{equation}
 \begin{split}
H_{j}=
 & J_{0}(S^{z}_{1}S^{z}_{2}+S^{z}_{4}S^{z}_{5})+
J_{1}S^{z}_{3}(S^{z}_{1}+S^{z}_{5})\\
&+J_{2}S^{z}_{3}(S^{z}_{2}+S^{z}_{4}).
\label{eq:H-ising-plaquette}
\end{split}
\end{equation}
It is easy to identify collinear ground states corresponding to Hamiltonian (\ref{eq:H-ising-plaquette}). 
In Fig. \ref{fig:ising-picture} we show the Ising states corresponding to $m=1/5$ and $m=3/5$ on the plaquette, where red(black) lines correspond to antiferromagnetic(ferromagnetic) couplings. 
Notice that, as the $J_1$ and $J_2$ are alternate in the material, two different minimal energy patterns corresponding
to $m=1/5$ can be found. 

As the 5 sites cells are connected via ferromagnetic couplings is straightforward to extend these local magnetic structures to the complete Kagom\'e strip lattice.
For $m=3/5$, the result of this interaction is a product state of individual plaquettes in the same state, 
and the (normalized) magnetization  is still $m=3/5$, as showed in Fig. \ref{fig:ising-picture} (Bottom).
For $m=1/5$ an alternated cell configurations can be build as showed in Fig. \ref{fig:ising-picture} (Top)

The signs in the $\langle S^z_{i}\rangle$ profile obtained by DMRG calculations are consistent with the classical picture. However quantum fluctuations contribute to lowering the strength of $S^z$.

It is expected that magnetic plateaux observed at $T=0$ disappear with increasing temperature. In order to estimate the temperature range where they could be observed we studied small clusters. From the previous analysis, we can expect that at low temperatures, the correlation length is shorter than the size of the unit cell. This allows estimating equilibrium magnetic properties, like magnetization by looking at equilibrium states of small subsystems. To carry it out, a reduced model of a single full unit cell with periodic conditions was considered. The magnetization as a function of the magnetic field and temperature was evaluated assuming a thermal equilibrium state
$$
m=\frac{M}{M_S} =\frac{\langle {\bf S}^z_{T}\rangle}{5} = \frac{{\rm Tr}{\bf S}_{T}^z e^{-{\bf H}_s /kT}}{5 {\rm Tr}e^{-{\bf H}_s /kT}}
$$
with ${\bf S}_{T}^z=\sum_i \vec{\bf S}_{i}^z$ the $z-$component of the total spin of the unit cell
{ and ${\bf H}_S$ the Hamiltonian of the reduced cell with periodic boundary conditions}.

\begin{figure}[t!]
  \centering
  \includegraphics[width=1.\columnwidth]{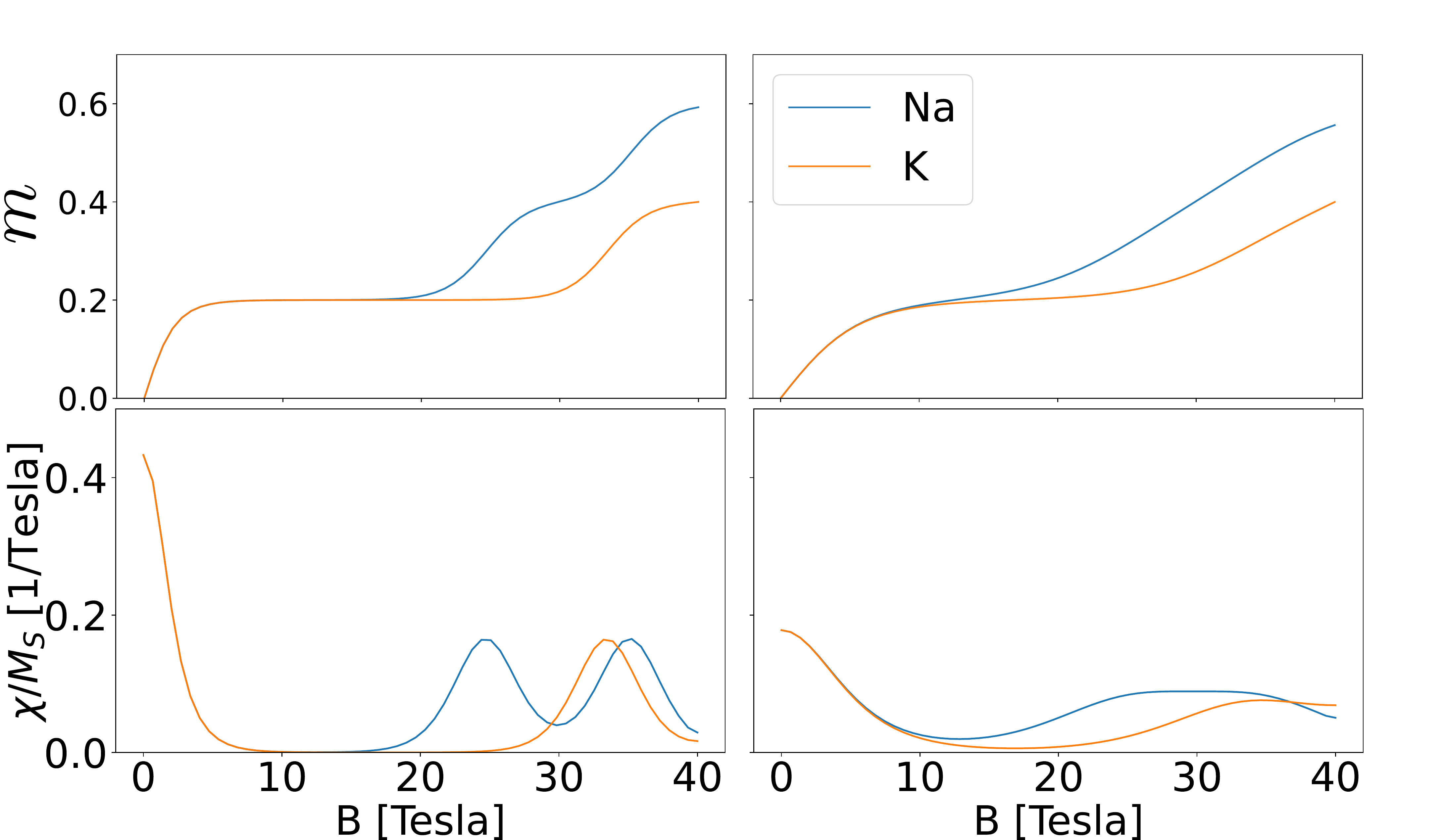}
  \caption{Magnetization (top) and susceptibility (bottom) for the Heisenberg model as a function of $B$ for $T=2 {\rm K}$ (left) and $T=5 {\rm K}$ (right) for Na-TS and K-TS compounds for couplings obtained for $U=5\rm{eV}$.}
  \label{fig:magTfinita}
\end{figure}

The results suggest that the G.S. plateaux at $m=1/5$ and $m=3/5$ disappear for temperatures around $5{\rm K}$, but are still manifested in the susceptibility.
Magnetization curves for Na-TS and K-TS are depicted for fixed, lower temperatures in Figure \ref{fig:magTfinita}.

% 
% {\color{red} Energía mínima en cada sector de magnetización a campo 0, en Kelvin:
%   \begin{tabular}{cc}
%     sector & energia \\
%     $\pm 1/5$ & $0$\\
%     $0$    & $0$\\
%     $\pm 2/5$ & $33.19$ \\
%     $\pm 3/5$ & $80.19$\\
%     $\pm 4/5$ & $202.79$\\
%     $\pm 5/5$ & $346.93$
%   \end{tabular}
% Notar que $\mu_B/k_B\approx 1.34 {\rm K}/{\rm T}$.
% }          

%

\section{Conclusions}
\label{sec:conclusions}

We have studied by a combination of DFT+$U$, full diagonalization and DMRG calculations two recently reported tellurite-sulfates, Na$_2$ Cu$_5$(TeO$_3$)(SO$_4$)$_3$(OH)$_4$ and K$_2$Cu$_5$(TeO$_3$)(SO$_4$)$_3$(OH)$_4$ (Na-TS and K-TS, respectively) that exhibit a 1D kagomé strip lattice.  
Our DFT+U calculations, performed as a function of the Hubbard term U, correctly predict the equilibrium structure of Na-TS and K-TS, which are irrespective of the spin configuration of both compounds. 
Based on an effective spin model, the five main magnetic exchange couplings of the distorted Kagom\'e strip lattice of Na-TS and K-TS were determined. Our calculations show that the relative couplings are nearly independent of the $U$ parameter. From these couplings, the Weiss temperature of Na-TS and K-TS were obtained.
Regarding magnetic properties of Na-TS and K-TS, both materials exhibit a mixture of ferromagnetic and antiferromagnetic couplings. The ferromagnetic ($J_u$) coupling is associated with the shortest Cu-Cu distance and a 90 degrees-like superexchange configuration for both compounds. The mixture was already speculated when synthesizing these materials in the framework of the Goodenough-Kanamori-Anderson rules.
Our calculations confirm these speculations and provide concise numerical values for the couplings and the corresponding uncertainties. These uncertainties turn out to be small enough to be considered small corrections on the effective model.
Finally, numerical calculations at $T = 0$ based on the obtained coupling constants reveals the existence of two plateaux in the magnetization curve.

\section*{Acknowledgments}
ILB, JMM, CAL, LE, AGR, VF. acknowledge support from CONICET.
This work was partially supported by CONICET, Argentina (grant no PIP 11220200102332CO, PIP 11220200101877CO, PIP 11220200101460CO and PIP 0039-2017) 
and UNLP (grants no.11/X845 and X896/20).

\vspace{1cm}

%.\bibliographystyle{unsrt}
%\bibliography{biblio}
%\bibliography{Referencias_exact_gs}

\bibliography{referencias}

%%%%%%%%%%%%%%%%%%%%%%%%%%%%%%%%%%%%%%%%%%%%%%%%%%%%%%%%%%%%%%%%%%%%%%%%%%%%%%%%
\end{document}